\documentclass[showpacs,floatfix,eqsecnum,aps,nofootinbib]{revtex4}
\usepackage{epsfig}

\begin{document}

\title{O(10) kinks: clash of symmetries on the brane and the gauge hierarchy problem.}
\author{Edward M. Shin}
\affiliation{School of Physics, Research Centre for High Energy
Physics, The University of Melbourne, Victoria 3010, Australia}
\author{Raymond R. Volkas}\email{r.volkas@physics.unimelb.edu.au}
\affiliation{School of Physics, Research Centre for High Energy
Physics, The University of Melbourne, Victoria 3010, Australia}

\begin{abstract}
We study kink or domain wall solutions in $O(10)$ Higgs models in the context of the
``clash of symmetries'' mechanism developed by Davidson, Toner, Volkas and Wali and, 
independently, by Pogosian and Vachaspati. We show that kink configurations employing 
Higgs fields in the ${\bf 45}$ (the adjoint representation) of $O(10)$ break up into 
three classes: those that at finite distances from the wall respect a $U(5)$ subgroup 
of $SO(10)$, and two others that respect the smaller subgroups $U(3) \otimes U(2)$ and 
$U(4) \otimes U(1)$. These smaller subgroups arise from the clash of symmetries mechanism: 
they are the intersections of two differently embedded $U(5)$ subgroups of $SO(10)$, the 
latter being the symmetries respected in asymptotia on opposite sides of the domain wall. 
The existence of the $SO(10) \to U(3) \otimes U(2) = SU(3) \otimes SU(2) \otimes U(1) 
\otimes U(1)' \equiv G_{SM} \otimes U(1)'$ class advances the search for a realistic 
brane world model wherein some of the required symmetry breaking is achieved via the clash 
of symmetries rather than the conventional mechanism. At the centres of the walls, the 
unbroken symmetries are enhanced. In the $U(3) \otimes U(2)$ case, the symmetry is $O(6) 
\otimes U(2)$, which is closely related to the Pati-Salam-like $SU(4) \otimes SU(2) \otimes 
U(1)$ group. If our universe is a brane located at the centre of such a wall, then we see the 
$O(10)$ symmetry as being strongly broken to $SU(4) \otimes SU(2) \otimes U(1)$. 
Interestingly, if the brane-world degrees of freedom enjoy a slight leakage off the wall, 
then an additional symmetry breakdown to $U(3) \otimes U(2) = G_{SM} \otimes U(1)'$ is 
effectively induced on the brane. This provides a possible framework within which to address at
least part of a gauge hierarchy problem: $O(10)$ is strongly broken to $SU(4) \otimes SU(2) 
\otimes U(1)$, then more weakly to $G_{SM} \otimes U(1)'$ depending on the amount of leakage 
off the brane. We also comment on kinks employing the ${\bf 10}$ and ${\bf 54}$ of $O(10)$.
\end{abstract}


\maketitle

\section{Introduction}

Soliton solutions of field theories relevant to particle physics, cosmology
and condensed matter physics
continue to fascinate \cite{soliton}. 
We shall study kinks or domain walls in certain $O(10)$
Higgs models in this paper. One motivation is the symmetry breaking
mechanism called the ``clash of symmetries'' proposed for brane-world models 
by Davidson, Toner, Volkas and Wali \cite{davidson}, and independently studied by Pogosian,
Vachaspati and collaborators for different reasons \cite{pg1,pg}. The symmetry $O(10)$ is
considered simply because $SO(10)$ is a possible grand unification group. Much of this
work is based on Ref.\ \cite{edward}.

We show that a certain kink arising in an $O(10)$-adjoint Higgs model
respects a $U(3) \otimes U(2)$ subgroup of $O(10)$ at all finite distances from 
the wall. At the centre of the wall, the unbroken symmetry is the larger
subgroup $O(6) \otimes U(2)$. This symmetry structure is interesting for model-building,
because $U(3) \otimes U(2)$ is isomorphic to 
$G_{SM} \otimes U(1)'$, where $G_{SM} = SU(3) \otimes SU(2)
\otimes U(1)$ is the standard model gauge group. Furthermore, the connected part of $O(6)$
is $SO(6) = SU(4)$ which means that a Pati-Salam-like group \cite{ps} 
is exact at the centre of the
wall. If our universe is a brane located at the centre of such a wall, the perpendicular
coordinate to which is an extra dimension, then we see the 
$O(10)$ symmetry as being strongly broken to $SU(4) \otimes SU(2) \otimes U(1)$; 
and, if the brane-world degrees of freedom enjoy a slight leakage off the
wall, then an additional symmetry breakdown to $U(3) \otimes U(2) = 
G_{SM} \otimes U(1)'$ is effectively
induced on the brane. This provides a possible framework within which to address at
least part of the gauge hierarchy problem: $O(10)$ is strongly broken to 
$SU(4) \otimes SU(2) \otimes U(1)$, then more weakly to $G_{SM} \otimes U(1)'$ depending
on the amount of leakage off the brane.\footnote{Since we are not considering gravity
yet, we will not address the Planck/electroweak hierarchy issue. Note also that
the specific kink we shall study in this paper leaves the electroweak symmetry
unbroken. The point we want to emphasise is that kinks of this nature have
something to say about ratios of symmetry breaking scales; they
provide a new {\it framework} for thinking about gauge hierarchies. It is then a 
model-building challenge to find a fully realistic embodiment of the general idea.}

These results exemplify our hope for what aid the model-builder might
receive from the clash of symmetries mechanism. Model-builders are regularly frustrated
that the beautiful enhanced symmetries they like to entertain, such as $SO(10)$,
apparently must be embedded in often unsightly models involving several Higgs 
fields and many free parameters. If only our Higgs fields could do more
work for us! Well, by admitting spatially varying Higgs field configurations
such as kinks into the game, it is clear that spatially varying symmetry
breaking patterns can be achieved, and perhaps even utilised in realistic models.
With no evidence for domain walls and the like in our Hubble volume, one must
turn to the brane-world hypothesis \cite{brane}
for help. It remains to be seen if the price
of extra dimensions is worth paying.

We would also like to remark that
apart from any specific applications, the symmetries of kinks
are also of interest
for their own sake, the deeper appreciation of the
variety of soliton solutions to non-trivial non-linear classical field theories
being a worthy goal.

Topologically stable domains walls arise when a discrete symmetry is spontaneously
broken. The vacuum manifold consists of disconnected pieces, and the kink 
configurations are classical solutions to the Euler-Lagrange equations that 
interpolate between two states from these disconnected pieces. We shall consider
theories with elementary scalar (synonymously Higgs) 
fields only in this paper. On one side of the
wall, the scalar fields asymptote to one of the possible vacua, while on the other
side they approach a different vacuum state, not continuously connected to
the former via any continuous symmetry transformation. In the simplest model
scalar field theories having topologically stable kink solutions, the vacuum
``manifold'' is simply a collection of disconnected points related to each other
by the spontaneously broken discrete transformations.

Models of this class have been well-studied. But, more recently, interest has been
growing in a richer class of models that contain spontaneously broken
continuous symmetries in addition to the discrete symmetry \cite{davidson,pg1,pg}.
The global minima of the Higgs potential generally break both the continuous
and the discrete symmetries spontaneously, and the disconnected pieces of the
vacuum manifold are no longer simply points but are actually manifolds 
formed from continua of degenerate vacua. Consider kink
configurations that interpolate between disconnected vacua in such a model.
In each topological class there is a continuum of solutions corresponding to
different choices for asymptotic vacua, and in general they have different energies.
Toplogical arguments guarantee that there is a stable kink within each class, the
one with lowest energy, but one has to do more analysis to actually identify which kink
that is.

Kinks within a class can have importantly different symmetry properties through
the ``clash of symmetries'' phenomenon. Suppose that the continuous symmetry
$G$ spontaneously breaks to the subgroup $H$ (accompanied by the simultaneous
breaking of a discrete symmetry). Each connected piece from the vacuum manifold
is given by the coset space $G/H$, replicated a discrete number of times
depending on the discrete symmetry breaking pattern. The clash of symmetries
corresponds to {\it differently embedded}, though isomorphic, subgroups $H(-\infty)$
and $H(+\infty)$ being respected asymptotically on opposite sides of the
wall. At non-asymptotic points, the unbroken symmetry is generally smaller than
$H$. We shall restrict our analysis to kinks respecting precisely the
intersection group, $H(-\infty) \cap H(+\infty)$, at all locations
a finite distance from the wall. These kinks are
spatially-dependent linear combinations of the asymptotic vacua.

While we shall focus on the adjoint representation, we shall also
briefly comment on $O(10)$ kinks using the vector 
and symmetric rank-2 tensor representations, the ${\bf 10}$ and ${\bf 54}$
respectively, purely out of general interest. 

Only topological kinks will be examined in this paper. Non-topological
kink configurations are also of general interest, but because their stability
is not {\it a priori} guaranteed they are of limited use for brane-world
models. One should note, however, that a rich collection of non-topological
kink solutions are expected to exist for the models we consider.

The rest of this paper is structured as follows. In Sec.\ \ref{45kink}
we discuss kink configurations generated by the ${\bf 45}$ of $O(10)$.
The symmetry structure is explained and analytically exact kink 
configurations are found for a particular slice through Higgs potential
parameter space.
Section \ref{otherkinks} briefly discusses the group theory of kinks
constructed from the ${\bf 10}$ and ${\bf 54}$ of $O(10)$, while
Sec.\ \ref{conclusion} is a conclusion.

\section{Kinks from the ${\bf 45}$ of $O(10)$}
\label{45kink}

\subsection{Symmetry breaking patterns and kink boundary conditions}

The representation space of the ${\bf 45}$ (adjoint representation) 
of $O(10)$ can be taken to
consist of real antisymmetric $10 \times 10$ matrices. Consider a
multiplet of Higgs bosons, $\Phi$, in this representation:
\begin{equation}
\Phi = (\phi_{ij}),\quad \phi_{ij} = - \phi_{ji}\,\, \in\,\, \Re,\quad 
i,j = 1,\ldots,10.
\label{eq:Phi}
\end{equation}
Its transformation law is
\begin{equation}
\Phi \to A \Phi A^T
\label{eq:PhiXfm}
\end{equation}
where $A$ is a real $10 \times 10$ orthogonal matrix. The connected part
of $O(10)$ is the $SO(10)$ subgroup formed from all determinant one (special), 
orthogonal, real $10 \times 10$ matrices. The det$(A)=-1$ (antispecial) subset of
$O(10)$ is disconnected from the $SO(10)$
submanifold. The two subspaces are related by a discrete $Z_2$ 
transformation induced by some suitably chosen antispecial element of $O(10)$.
For instance, the $O(10)$ matrices
\begin{equation}
{\bf 1}_{10}\quad {\rm and}\quad Z \equiv {\rm diag}(\sigma_1,\sigma_1,\sigma_1,
\sigma_1,\sigma_1),
\label{eq:Z}
\end{equation}
where $\sigma_1$ is the first of the Pauli matrices $\sigma_{1,2,3}$
and ${\bf 1}_N$ is the $N \times N$ identity, form a $Z_2$
subgroup that can be used for this purpose. It suffices to note that
any antispecial $O(10)$ matrix can be multiplied by $Z$ to produce an $SO(10)$
matrix, and vice-versa. These remarks will be relevant later for the topological
stability question.

We shall use quartic Higgs potentials in this paper because of their familiarity,
though we are well aware that this restriction may not be necessary for brane-world
models and other contexts. Lacking, as yet, criteria for choosing one
non-quartic potential over another, we feel it is sensible to study the quartic
case first.

The most general $O(10)$ invariant quartic potential for $\Phi$ is
\begin{equation}
V = \frac{1}{2}\, \mu^2\, {\rm Tr}(\Phi^2) + \frac{1}{4}\, \lambda_1\, 
{\rm Tr}(\Phi^2)^2 + \frac{1}{4}\, \lambda_2\, {\rm Tr}(\Phi^4),
\label{eq:V}
\end{equation}
where we take $\mu^2 > 0$, having noted that Tr$(\Phi^2) = -\sum(\phi_{ij})^2$.
The conventionally normalised kinetic energy term is
\begin{equation}
T = - \frac{1}{4} {\rm Tr}(\partial^{\mu} \Phi\, \partial_{\mu} \Phi),
\label{eq:T}
\end{equation}
with Minkowski signature $(+,-,-,-)$.
The prefactor is $1/4$ rather than $1/2$ because each independent component of
$\Phi$ occurs twice in the summation Tr$(\Phi^2)$.

Notice that $V$ and $T$ are invariant under the discrete $Z_2$ symmetry defined by
\begin{equation}
\Phi \to -\Phi.
\label{eq:Z2}
\end{equation}
This accidental discrete symmetry arises because the $O(10)$ invariant cubic
term Tr$(\Phi^3)$ vanishes identically. This symmetry is also outside $O(10)$,
being distinct from the $Z_2$ relating the special and antispecial subsets
of $O(10)$. This is easily established. Consider an $O(10)$ invariant theory
with three independent ${\bf 45}$'s, denoted by $\Phi_{1,2,3}$. The cubic
term Tr$(\Phi_1\, \Phi_2\, \Phi_3)$ is nonzero and respects $O(10)$ but not
$\Phi_{1,2,3} \to -\Phi_{1,2,3}$.

The symmetry breaking patterns induced by Eq.\ (\ref{eq:V}) were deduced by L.-F.\ Li
in his classic study of spontaneous symmetry breaking \cite{li}. 
The first step is to 
use the theorem from linear algebra establishing that any antisymmetric $2n \times 2n$
real matrix can be transformed as per Eq.\ (\ref{eq:PhiXfm}) to the ``standard form''
\begin{equation}
\Phi = {\rm diag}(a_1\, \epsilon\, ,\, a_2\, \epsilon\, ,\, \ldots\, ,\, a_n\, \epsilon),
\label{eq:standardform}
\end{equation}
where the $a_i$ are real numbers and
\begin{equation}
\epsilon \equiv i\, \sigma_2 = \left( \begin{array}{cc}
0 & 1 \\ -1 & 0 \end{array} \right).
\end{equation}
In this basis,
\begin{equation}
V = -\mu^2\, \sum_{i=1}^{5} a_i^2 + \lambda_1\, \left( \sum_{i=1}^{5} a_i^2 \right)^2
+ \frac{1}{2}\, \lambda_2\, \sum_{i=1}^{5} a_i^4,
\label{eq:rewrittenV}
\end{equation}
where we restrict our attention to constant fields only, of course.
The potential
is bounded from below in the $\lambda_{1,2}$ region defined by
\begin{equation}
10\lambda_1 + \lambda_2 > 0\ \ {\rm for}\ \ \lambda_2 > 0\, ;\qquad
2\lambda_1 + \lambda_2 > 0\ \ {\rm otherwise}.
\end{equation}
Straightforward algebra establishes that for $\lambda_2 > 0$, the 
global minima of $V$ are at
\begin{equation}
a_i^2 = \frac{\mu^2}{10\lambda_1 + \lambda_2}\quad \forall i,
\label{eq:globalminU5}
\end{equation}
while for $\lambda_2 < 0$ they are at
\begin{equation}
a_1^2 = \frac{\mu^2}{2\lambda_1 + \lambda_2},\quad a_2 = a_3 = a_4 = a_5 = 0,
\label{eq:globalminO8xSO2}
\end{equation}
and permutations of this pattern.

Using $M\, \epsilon\, M^T = (\det M)\, \epsilon$ (where $M$ is any $2\times 2$ matrix) 
it is obvious by inspection
that Eq.\ (\ref{eq:globalminO8xSO2}) corresponds to the breakdown
\begin{equation}
O(10) \to O(8) \otimes SO(2).
\end{equation}
For the specific global minimum displayed in Eq.\ (\ref{eq:globalminO8xSO2}),
the unbroken $SO(2)$ acts on the subspace defined by the upper-left $2\times 2$
block of $\Phi$. The permutations correspond to moving the $SO(2)$ invariant
subspace along the diagonal. The $O(8)$ invariant subspace moves in concert. 
These possibilities define what we shall call ``different embeddings of the 
$O(8) \otimes SO(2)$ subgroup in $O(10)$''. Clearly, if we depart from the
standard form $\Phi$ of Eq.\ \ref{eq:standardform} we discover a continuum of
different embeddings. However, as we shall show later on, 
the kink solutions of the Euler-Lagrange
equations for our $SO(10)$ adjoint-Higgs model {\it must} assume the standard form (with
respect to some basis). Thus no generality is lost, in the kink case at least, 
by working with the
finite number of discretely-different embeddings consistent with the standard 
form, and this is what we shall mean by the term ``different embeddings''. The embeddings
are physically equivalent for global homogeneous vacua, but not for
inhomogeneous configurations such as kinks.

Our main interest in this paper rests with the global minima of Eq.\ (\ref{eq:globalminU5}).
They are invariant under $U(5)$
subgroups of $O(10)$, the derivation of which we now review. 
To understand how a $U(5)$ sits inside $O(10)$, one needs
to generalise the well-known $U(1) \leftrightarrow SO(2)$ mapping (isomorphism 
actually), given simply by
\begin{equation}
e^{i\theta} \leftrightarrow \left( \begin{array}{cc} \cos\theta & -\sin\theta \\
\sin\theta & \cos\theta \end{array} \right).
\end{equation}
Let $h$ be the mapping of any complex number $re^{i\theta}$ to the corresponding
$SO(2)$ matrix multiplied by the positive (or zero) real number $r$. 
The image of $h$, Im$(h)$, equals $[0,+\infty) \otimes SO(2)$. Observe that
$h$ preserves additive as well as multiplicative structure.

An obvious generalisation of $h$ is to map $n \times n$ matrices of
complex numbers to $2n \times 2n$ matrices of real numbers where
the latter may be thought of as $n\times n$ matrices with $2\times 2$
matrix entries, with the $2\times 2$ blocks just being the $h$-images of the 
original complex number entries. We shall avoid pedantry
by also calling this mapping $h$. 

Observe that
\begin{equation}
h(M^{\dagger}) = h(M)^T
\label{eq:daggertotranspose}
\end{equation}
for any matrix $M$. Let $U \in U(5)$. Then
\begin{equation}
h(U^{\dagger}\, U)=h({\bf 1}_5) = {\bf 1}_{10}.
\label{eq:identityhmap}
\end{equation}
But, because $h$ acting on complex numbers preserves {\it both} additive and 
multiplicative structure, it follows that
\begin{equation}
h(U^{\dagger}\, U) = h(U^{\dagger})\, h(U) = h(U)^T\, h(U),
\label{eq:unitaryhmap}
\end{equation}
where the last equality uses Eq.\ (\ref{eq:daggertotranspose}).
Similarly,
\begin{equation}
h(U_1\, U_2) = h(U_1)\, h(U_2)
\label{eq:hmatrixmultpreserving}
\end{equation}
for any $U_{1,2} \in U(5)$.
Combining Eqs.\ (\ref{eq:identityhmap}-\ref{eq:hmatrixmultpreserving}), we see
that $h$ maps $U(5)$ into $O(10)$. Actually, since $U(5)$ is a connected manifold,
$h$ maps $U(5)$ into $SO(10)$. 

Return now to the global minima of Eq.\ (\ref{eq:globalminU5}) and consider the
specific one given by
\begin{equation}
\langle\Phi\rangle_1 \equiv a_{\rm min}\, {\rm diag}(\epsilon\, ,\, \epsilon\, ,\, 
\epsilon\, ,\, \epsilon\, ,\, \epsilon),
\label{eq:Phivac1}
\end{equation}
where $a_{\rm min} \equiv \sqrt{\mu^2/(10\lambda_1 + \lambda_2)}$. Observing
that $\epsilon = h(-i)$, we see that
\begin{eqnarray}
\langle\Phi\rangle_1 & = & a_{\rm min}\, h(-i\, {\bf 1}_5)\nonumber\\
& = & a_{\rm min}\, h( U\, (-i\, {\bf 1}_{5})\, U^{\dagger})\nonumber\\
& = & a_{\rm min}\, h(U)\, h(-i\, {\bf 1}_5)\, h(U^{\dagger})\nonumber\\
& = & h(U)\, \langle\Phi\rangle_1\, h(U)^T.
\label{eq:U5invPhi}
\end{eqnarray}
That is, $\langle\Phi\rangle_1$ is invariant under the $U(5)$ subgroup
defined through the mapping $h$. Since $U(5)$ is a maximal subgroup
of $SO(10)$, it exhausts the invariances of $\langle\Phi\rangle_1$. 

The other global minima defined by Eq.\ (\ref{eq:globalminU5}) are obtained
from Eq.\ (\ref{eq:Phivac1}) by writing down all possible combinations of
$+\epsilon$ and $-\epsilon$ down the diagonal. They partition into
two classes, with the members of a given class related by $SO(10)$
transformations. The two classes are related to each other through
an overall change of sign, as per the discrete transformation of 
Eq.\ (\ref{eq:Z2}). To prove these assertions, we make use of the
$SO(10)$ matrix
\begin{equation}
{\rm diag}({\bf 1}_2\, ,\, {\bf 1}_2\, ,\, {\bf 1}_2\, ,\, \sigma_1\, ,\, \sigma_1)
\label{eq:type1}
\end{equation}
and permutations thereof (type 1), together with the $SO(10)$ matrix
\begin{equation}
{\rm diag}({\bf 1}_2\, ,\, \sigma_1\, ,\, \sigma_1\, ,\, \sigma_1\, ,\, \sigma_1)
\label{eq:type2}
\end{equation}
plus its permutations (type 2). Noting that $\sigma_1\, \epsilon\, \sigma_1^T = -\epsilon$,
we see that the type 1 matrices reverse the signs of two of the $\epsilon$'s
placed along the diagonals of the various $\langle\Phi\rangle$'s, while type
2 matrices reverse the signs of four $\epsilon$'s. 
Matrices with an odd number of $\sigma_1$'s along the diagonal, thus capable of
reversing the signs of an odd number of $\epsilon$'s, are antispecial.

Denoting each global vacuum using the obvious notation
$a_{\rm min}{\rm diag}(\epsilon\, ,\, \epsilon\, ,\, \epsilon\, ,\, 
\epsilon\, ,\, \epsilon)$ $\to$ $(+,+,+,+,+)$ and so on, it is clear
that the two classes of vacua are:
\begin{equation}
(+,+,+,+,+),\quad (+,+,+,-,-)\, \&\, {\rm perms.},\quad (+,-,-,-,-)\, 
\&\, {\rm perms.},\quad ({\it class\ 1})
\label{eq:class1}
\end{equation}
and
\begin{equation}
(-,-,-,-,-),\quad (-,-,-,+,+)\, \&\, {\rm perms.},\quad (-,+,+,+,+)\, 
\&\, {\rm perms.}\quad ({\it class\ 2})
\label{eq:class2}
\end{equation}
Stated another way, the situation is the following: The set of $SO(10)$
transforms of $(+,+,+,+,+)$ (or any other class 1 vacuum), generates
a connected piece of the vacuum manifold in which there are particular
points corresponding to the class 1 configurations. They are precisely 
those that also assume the standard form of Eq.\ (\ref{eq:standardform}).
There is a second connected piece consisting of the $SO(10)$ transforms
of $(-,-,-,-,-)$, within which appear the class 2 minima.
The two pieces are related by the spontaneously
broken discrete symmetry $\Phi \to -\Phi$.\footnote{There is a nicety
worth mentioning here to forestall possible confusion. 
The $Z_2$ transformation of 
Eq.\ (\ref{eq:Z2}) is outside of $O(10)$. However, {\it when $\Phi$'s
of standard form only} are considered, the antispecial $O(10)$ matrix
${\rm diag}(\sigma_1\, ,\, \sigma_1\, ,\, \sigma_1\, ,\, \sigma_1\, ,\, 
\sigma_1)$ also induces $\Phi \to -\Phi$. Technically, it appears that
it is the diagonal subgroup of these two $Z_2$'s that is spontaneously
broken. The important fact, though, is that the class 1 and class 2
vacua are disconnected from each other, whatever discrete symmetry
you blame it on, so topological stability
for some kink is assured. We have checked numerically that there is no
{\it special} orthogonal matrix that also induces $\Phi \to -\Phi$
for standard form $\Phi$'s.}

Topologically stable kink configurations interpolate between a
class 1 global minimum and a class 2 global minimum. Let $z$ be
the spatial coordinate perpendicular to the domain wall. Without
loss of generality, we may take the boundary condition at $z=-\infty$ 
to be a particular class 2 vacuum, say
\begin{equation}
\Phi(-\infty) = - a_{\rm min}\, {\rm diag}
(\epsilon\, ,\,\epsilon\, ,\, \epsilon\, ,\,\epsilon\, ,\,\epsilon).
\label{eq:minusbc}
\end{equation}
All other choices for the class 2 vacuum can be obtained by suitably 
transforming the whole kink, and $z \to -z$ trivially interchanges the
roles of class 1 and 2 minima. But different choices of class 1 vacua
for the boundary condition at $z = +\infty$ correspond to physically
distinct kinks, though all lie within the same topological class. The three
nontrivially different choices are
\begin{equation}
\Phi(+\infty) = \left\{  \begin{array}{ccc}
\Phi_{\rm min}^{(5)} & \equiv & a_{\rm min}\, {\rm diag}
(\epsilon\, ,\, \epsilon\, ,\, \epsilon\, ,\, \epsilon\, ,\, \epsilon) \\
\Phi_{\rm min}^{(3,2)} & \equiv &  a_{\rm min}\, {\rm diag}
(\epsilon\, ,\, \epsilon\, ,\, \epsilon\, ,\, - \epsilon\, ,\, - \epsilon) \\
\Phi_{\rm min}^{(4,1)} & \equiv & a_{\rm min}\, {\rm diag}
(\epsilon\, ,\, - \epsilon\, ,\, - \epsilon\, ,\, - \epsilon\, ,\, - \epsilon)
\end{array} \right. .
\label{eq:plusbc}
\end{equation}
Permutations of the minus signs in the last two of these vacua correspond to
kink configurations that are easily derived from those obeying the above, so we
need not consider them explicitly. The superscripts $(5)$, $(3,2)$ and $(4,1)$
label the symmetry unbroken by the kink at finite distances from the wall, respectively
\begin{equation}
U(5),\qquad U(3)\otimes U(2)\quad {\rm and}\qquad U(4)\otimes U(1),
\end{equation}
as we now explain.

A natural ansatz for kink configurations that interpolate between the stated boundary 
conditions is
\begin{equation}
\Phi_k(z) = \alpha(z)\, \Phi(-\infty) + \beta(z)\, \Phi(+\infty),
\label{eq:alphabeta}
\end{equation}
where
\begin{equation}
\alpha(-\infty) = 1,\quad \alpha(+\infty) = 0,\quad
\beta(-\infty) = 0,\quad \beta(+\infty) = 1.
\label{eq:alphabetabcs}
\end{equation}
Such kinks are a subset of configurations that maintain standard form [as
defined by Eq.\ (\ref{eq:standardform})] over all space. 
We shall see in the next section that although a general standard-form
$\Phi$ has five independent functions $a_i$, the kinks have only two (called $\alpha$ and
$\beta$ above). The maintenance of standard form for all $z$ will be justified in
the next subsection.

When $\Phi(+\infty) = \Phi_{\rm min}^{(5)}$, the configuration
is proportional to ${\rm diag}(\epsilon\, ,\, \epsilon\, ,\, \epsilon\, ,\, 
\epsilon\, ,\, \epsilon)$ for all $z$, so the unbroken symmetry remains the 
$U(5)$ subgroup defined by the $h$ map [denote this $U(5)$ by $U(5)_h$].
As we discuss later, the configuration has vanishing fields when $z=0$, so
the $O(10)$ symmetry is completely restored at the centre of the wall.
We shall call a configuration obeying these boundary conditions a 
``symmetric kink'', as there is no clash of symmetries.

When $\Phi(+\infty) = \Phi_{\rm min}^{(3,2)}$, the configuration
is {\it not} proportional to ${\rm diag}(\epsilon\, ,\, \epsilon\, ,\, \epsilon\, ,\, 
\epsilon\, ,\, \epsilon)$ for any $|z| < \infty$. 
At $z = -\infty$, the unbroken
symmetry is $U(5)_h$, while at $z= +\infty$ it is a differently
embedded $U(5)$, related to the former via conjugation using a type 1 
$SO(10)$ transformation defined in Eq.\ (\ref{eq:type1}). Symbolically,
\begin{equation}
U(5)_{1} = Q_1\, U(5)_{h}\, Q_1^T
\label{eq:U5one}
\end{equation}
where $Q_1 = {\rm diag}({\bf 1}_2\, ,\, {\bf 1}_2\, ,\, {\bf 1}_2\, ,\, \sigma_1\, 
,\, \sigma_1)$. At finite values of $z$, the unbroken symmetry is the
intersection
\begin{equation}
U(5)_h \cap U(5)_1 = U(3) \otimes U(2),
\end{equation}
as established in the Appendix. The $U(3) \otimes U(2)$ invariance
group arises from the clash of symmetries mechanism. This is isomorphic to the
standard model gauge group with an additional Abelian factor, and offers hope
that the clash of symmetries mechanism could be employed within a
realistic brane-world model. As we show in the next section, at $z=0$ the
unbroken symmetry is the larger group $O(6) \otimes U(2)$. The application to
the gauge hierarchy problem foreshadowed in the introduction will be discussed
more fully later on. We dub such a configuration an ``asymmetric kink''.

Finally, when $\Phi(+\infty) = \Phi_{\rm min}^{(4,1)}$, the unbroken
symmetry at finite $z$ is
\begin{equation}
U(5)_h \cap U(5)_2 = U(4) \otimes U(1),
\end{equation}
where
\begin{equation}
U(5)_{2} = Q_2\, U(5)_{h}\, Q_2^T,
\end{equation}
with $Q_2 = {\rm diag}({\bf 1}_2\, ,\, \sigma_1\, ,\, \sigma_1\, ,\, \sigma_1\, 
,\, \sigma_1)$ being a type 2 $SO(10)$ matrix as per Eq.\ (\ref{eq:type2}).
At $z=0$ it increases to $U(4) \otimes O(2)$. These are ``super-asymmetric kinks''.

\subsection{Solving the Euler-Lagrange equations}

We are now almost ready to solve the Euler-Lagrange equations using the
ansatz of Eqs.\ (\ref{eq:alphabeta}) and (\ref{eq:alphabetabcs}) subject to
the boundary conditions of Eqs.\ (\ref{eq:minusbc}) and (\ref{eq:plusbc}).

But first, we need to discuss the justification for considering kink
configurations that keep to the standard form of Eq.\ (\ref{eq:standardform})
at all $z$. In Sec.\ III of Ref.\cite{pg1}, Pogosian and Vachaspati prove a
powerful analogous theorem for $SU(N)$-adjoint kinks: writing the
adjoint as a traceless $N\times N$ matrix, they show that 
it suffices to consider ansatze where this matrix remains diagonal for all $z$.
The analogue of diagonal form for $SO(10)$ adjoints is the standard form.
The Pogosian-Vachaspati argument is easily adapted to this case as we now
show.

Consider a putative kink configuration of the form
\begin{equation}
\Phi(z) = S(z) + N(z),
\end{equation}
where $S(z)$ is of standard form, and $N(z)$ is a perturbation
taking ``completely non-standard form''. Completely non-standard 
is the analogue here
of completely non-diagonal in the $SU(N)$ case: we define them to be
matrices with $2\times 2$ blocks of zero-matrices 
along the diagonal. [In addition, $N(z)$ must be antisymmetric of course.]
We substitute for $\Phi(z)$ in the energy density
and look for terms that are linear in $N$. If these are absent,
then we know that solutions cannot simultaneously have 
$S$ and $N$ contributions. By taking an asymptotic vacuum of standard
form, we then know that standard form is maintained for all 
$z$.\footnote{Configurations of completely non-standard form are
presumably $SO(10)$ transforms of standard form configurations.}

Examine the quadratic term in $V$ first:
\begin{equation}
{\rm Tr}(\Phi^2) = {\rm Tr}(S^2) + 2\, {\rm Tr}(SN) + {\rm Tr}(N^2).
\end{equation}
But it is easily verified by direct matrix multiplication that the
product of a standard matrix and a completely non-standard matrix is another
completely non-standard matrix, and hence the trace of it is zero: no term linear
in $N$ (or $S$ for that matter) is generated. This takes care
of the kinetic energy term as well as the ${\rm Tr}(\Phi^2)^2$
term in $V$. The potentially dangerous term in
${\rm Tr}(\Phi^4)$ is
\begin{equation}
{\rm Tr}[\, S^2\,(SN + NS) + (SN + NS)\, S^2\, ],
\end{equation}
which also vanishes identically for all $S$ and $N$. To see this
note that $S^2$ is a diagonal matrix (recall that
$\epsilon^2 = -{\bf 1}_2$), so each term in this trace is the product
of a diagonal and a completely non-standard matrix, and hence is a
completely non-standard 
matrix itself. We thus conclude that if we adopt the basis where one of the 
boundary conditions assumes
standard form, then the whole configuration is required to also have the
standard form. This powerful result greatly simplifies our analysis.
In particular, if we are able to find all standard-form kink solutions, then
we are guaranteed that the one with lowest energy will be topologically
stable.

Specialising to static configurations that depend just on a single
spatial coordinate $z$, the Euler-Lagrange equations yield
\begin{equation}
a_i'' = 2 \left[ -\mu^2 + 2 \lambda_1 \sum_{j=1}^{5} a_j^2 \right] a_i 
+ 2 \lambda_2 a_i^3,
\label{eq:generaldes}
\end{equation}
using Eqs.\ (\ref{eq:T}), (\ref{eq:standardform}) and (\ref{eq:rewrittenV}).
Observe that the $a_i$'s appear symmetrically in these differential equations, so only
the boundary conditions distinguish them. Furthermore, the equations are symmetric
under the parity inversion
\begin{equation}
z \to -z,
\label{eq:parity}
\end{equation}
and the boundary conditions for each $a_i$ are parity transforms of each other:
$a_i(-\infty) = \pm a_i(+\infty)$. This means that we can take
the solutions to be partitioned into odd and even parity classes.\footnote{Translational
invariance allows the centre of the domain wall to lie at any finite value of $z$.
Without loss of generality, we shall define this point to be $z=0$.} 
Furthermore, they 
must take the form of Eq.\ (\ref{eq:alphabeta}), and, as discussed in the Appendix, this
leads to the unbroken symmetries at $0 < |z| < \infty$ being the intersections of the
invariance groups of the asymptotic vacua.

When $\lambda_1 \neq 0$, Eqs.\ (\ref{eq:generaldes}) are coupled non-linear differential 
equations that can probably only
be solved numerically. For the purposes of this paper, we shall consider
the special point
\begin{equation}
\lambda_1 = 0,
\end{equation}
so that the equations decouple to
\begin{equation}
a_i'' =  2 \left( - \mu^2 
+ \lambda_2 a_i^2 \right) a_i,
\end{equation}
and analytical solutions are easily obtained. 

The symmetric kink obeys
\begin{equation}
a_1 = a_2 = a_3 = a_4 = a_5 \equiv f\quad {\rm with}\quad f(-\infty) = -a_{\rm min} = 
-\frac{\mu}{\sqrt{\lambda_2}},\quad f(+\infty) = +a_{\rm min},
\label{eq:symkinkansatz}
\end{equation}
where we define $\mu$ to be positive. The function $f$ satisfies
$f'' = 2f(-\mu^2 + \lambda_2 f^2)$, and the solution obeying the boundary
conditions in Eq.\ (\ref{eq:symkinkansatz}) is simply
\begin{equation}
f(z) = a_{\rm min} \tanh(\mu z),
\label{eq:fsoln}
\end{equation}
the ``archetypal'' kink-like function furnished by a quartic potential. Notice
that it is an odd function, so it vanishes at $z=0$, the centre of the wall.

Asymmetric kinks obey
\begin{equation}
a_1 = a_2 = a_3 \equiv f,\quad a_4 = a_5 \equiv g,
\end{equation}
with
\begin{equation}
f(-\infty) = -a_{\rm min},\quad f(+\infty) = +a_{\rm min}\quad
{\rm and}\quad g(-\infty) = g(+\infty) = - a_{\rm min},
\label{eq:fgbcs}
\end{equation}
which obey $f'' = 2f(-\mu^2 + \lambda_2 f^2)$ and $g'' = 2g(-\mu^2 + \lambda_2 g^2)$.
The solutions are 
\begin{equation}
f(z) = a_{\rm min} \tanh(\mu z),\quad g(z) = - a_{\rm min},
\label{eq:fgsolns}
\end{equation}
where $f$ is an odd function and $g$ is an even function. The ``$f$-block'' in the
matrix $\Phi$ vanishes at the centre of the wall.

Finally, super-asymmetric kinks conform to the ansatz
\begin{equation}
a_1 = f,\quad a_2 = a_3 = a_4 = a_5 = g,
\end{equation}
with the boundary conditions as per Eq.\ (\ref{eq:fgbcs}) and the same differential
equations as before satisfied. The solutions for $f$ and $g$ are as given 
in Eq.\ (\ref{eq:fgsolns}).

For $\lambda_1 \neq 0$, the functions $f$ and $g$ will depart form the forms
deduced above, but they must remain odd and even functions of $z$, respectively. 
This implies that symmetric kinks always vanish at $z=0$, asymmetric kinks
have a vanishing $6 \times 6$ block there, while super-asymmetric kinks have a 
vanishing $2 \times 2$ block. This immediately implies that the unbroken symmetry
at $z=0$ is the full $O(10)$ for symmetric kinks, $O(6) \otimes U(2)$ for asymmetric
kinks, and $U(4) \otimes O(2)$ for super-asymmetric kinks.

As explained in the previous subsection, the three different kink configurations
have distinct symmetry properties. In addition, they have different energy densities.
The Hamiltonian density for a $t,x,y$-independent $\Phi$ is
\begin{equation}
{\cal H} = -\frac{1}{4}\, {\rm Tr}(\Phi'\, \Phi') +
\frac{1}{2}\, \mu^2\, {\rm Tr}(\Phi^2) + \frac{1}{4}\, \lambda_1\, {\rm Tr}(\Phi^2)^2
+ \frac{1}{4}\, \lambda_2\, {\rm Tr}(\Phi^4) - V_0,
\label{eq:Ham}
\end{equation}
where 
\begin{equation}
V_0 = - \frac{5}{2} \frac{\mu^4}{10\lambda_1 + \lambda_2}
\end{equation}
is a constant chosen so that the potential energy minimum is at zero.
The surface energy density $\rho$ of a domain wall is then
\begin{equation}
\rho = \int_{-\infty}^{+\infty}\, {\cal H}\, dz.
\label{eq:rho}
\end{equation}
For standard-form configurations
\begin{equation}
{\cal H} = \frac{1}{2}\, \sum_{i=1}^{5}\, (a_i')^2 
- \mu^2\, \sum_{i=1}^{5} a_i^2 + \lambda_1\, \left( \sum_{i=1}^{5} a_i^2 \right)^2
+ \frac{1}{2}\, \lambda_2\, \sum_{i=1}^{5} a_i^4 - V_0.
\label{eq:rewrittenH}
\end{equation}
Each kink configuration has $n$ copies of $f$ and $5-n$ copies of $g$, where $n=5,3,1$
for symmetry, asymmetric and super-asymmetric kinks, respectively.
With $\lambda_1 = 0$, Eqs.\ (\ref{eq:fsoln}), (\ref{eq:rho})
and (\ref{eq:rewrittenH}) together with $g^2 = a^2_{\rm min} = \mu^2/\lambda_2$ yield
\begin{equation}
\rho(n) = \frac{4}{3}\, n \, \frac{\mu^3}{\lambda_2}.
\end{equation}
We therefore see that for the $\lambda_1 = 0$ case the super-asymmetric kink
has the lowest energy and is therefore topologically stable. The asymmetric kink
has the next lowest energy density, while the symmetric kink is the most energetic.
This pattern arises because there is an energy cost, equally contributed to by kinetic and 
potential energy, associated with the spatially varying function $f$. It is interesting
to note that the $\lambda_1$ term could conceivably re-order the energy hierarchy
of the kinks. Negative values of $\lambda_1$ are allowed, because any 
$\lambda_1 > - \lambda_2/10$ leads to a potential bounded from below. Because this
term is a trace squared, it has a quadratic rather than a linear dependence on $n$,
so it might be able to effect a re-ordering.

The important observation here is that {\it a kink configuration displaying the clash
of symmetries is energetically favoured in this case}. If one did not carefully
consider the different ways that $U(5)$ can be embedded in $SO(10)$, then one might
be prone to finding only the symmetric kink and falsely concluding that it was stable
for topological reasons. Note that from the point of view of our brane-world
motivation, quartic potentials with $\lambda_1 = 0$ appear to be unsuitable, because the
asymmetric kink is unstable. A full analysis of what potential or potentials would be 
suitable is well beyond the scope of this paper.

Table \ref{table} summarises the results for $O(10)$ adjoint kinks.

\begin{table*}
\caption{\label{table}A summary of kinks arising in the O(10) adjoint-Higgs model. 
See the main text for definitions of terms. The boundary conditions at $z=\pm\infty$
can be swapped.}
\begin{ruledtabular}
\begin{tabular}{c|c|c|c|c|c}
Name of kink & \multicolumn{4}{c|}{Unbroken subgroups at specified locations} & Energy density for $\lambda_1=0$.\\
\hline
& $z=-\infty$ & $z=+\infty$ & $0<|z|<\infty$ & z=0 & \\ 
\hline
Symmetric & $U(5)_h$ & $U(5)_h$ & $U(5)_h$ & $O(10)$ & $20\mu^3/3\lambda_2$\\
Asymmetric & $U(5)_h$ & $U(5)_1$ & $U(3)\otimes U(2)$ & $O(6)\otimes U(2)$ & $4\mu^3/\lambda_2$\\
Superasymmetric & $U(5)_h$ & $U(5)_2$ & $U(4)\otimes U(1)$ & $U(4)\otimes O(2)$ & 
$4\mu^3/3\lambda_2$\\
\end{tabular}
\end{ruledtabular}
\end{table*}

\subsection{Application to the gauge hierarchy problem}

We now expand on the possible application to the gauge hierarchy problem. Suppose
the coordinate $z$ defines an extra spatial dimension. Suppose further that our
universe is a brane located at $z=0$. It has degrees of freedom localised to it,
perhaps dynamically, but they interact with the Higgs field $\Phi$ which can
propagate into the extra dimension. Focus on the effective physics of the 
brane-confined degrees of freedom. The symmetry breaking induced by the $\Phi$-kink
is communicated through these interactions to the brane-world fields. In the 
Lagrangian, the brane-world fields lie in multiplets of $O(10)$. But the effective
theory will feature strong $O(10)$ breaking. We speculate, we think plausibly,
that the effective brane-world will, through leakage off the brane, be sensitive
to physical conditions at finite values of $z$ close to $z=0$. This means that
the effective theory will feature an additional symmetry breakdown to $U(3) \otimes
U(2) = G_{SM} \otimes U(1)'$, with the strength of this breaking given by the
amount of leakage off the brane. If the leakage is small, as one would expect it to
be, then the effective $O(6) \otimes U(2) \to G_{SM} \otimes U(1)'$ symmetry breaking
scale should be smaller than the effective $O(10) \to O(6) \otimes U(2)$ scale.
This relates the associated gauge hierarchy to both the symmetry structure of the
kink and the physics of leakage off the brane, providing a novel framework within
which to understand the origin of such a hierarchy.

Our specific proposal here provides a new schematic framework for addressing a
gauge hierarchy problem. It clearly requires additional nontrivial features to
become a realistic model: cogent reasons for having some degrees of freedom
confined to the brane while others propagate into the extra dimension, the
inclusion of gravity, a well-motivated potential inducing the correct kind
of kink-induced symmetry breaking (something like the asymmetric kink should be
topologically stable), and a way to induce electroweak symmetry and $U(1)'$
breakdown.

\subsection{Kinks from $O(10) \to O(8) \otimes SO(2)$.}

For completeness, we briefly discuss the clash of symmetry patterns that can be
obtained from the $O(10) \to O(8) \otimes SO(2)$ regime [see Eq.\ (\ref{eq:globalminO8xSO2})].
The global minima consist of a single $\epsilon$ matrix somewhere along the
diagonal, and zeroes everywhere else \cite{li}. Let the boundary condition at $z = - \infty$
be
\begin{equation}
\Phi(-\infty) = - a_{\rm min}\, {\rm diag}(\epsilon\, ,\, {\bf 0}_2\, ,\, {\bf 0}_2\, ,\, 
{\bf 0}_2\, ,\, {\bf 0}_2),
\end{equation}
where ${\bf 0}_n$ is the $n\times n$ matrix of zeroes. Arguments similar to that
in the Appendix reveal that:
\begin{eqnarray}
{\rm for}\ \Phi(+\infty) & = & a_{\rm min}\, {\rm diag}({\bf 0}_1\, ,\, \epsilon\, ,\, 
{\bf 0}_1\, ,\, {\bf 0}_2\, ,\, {\bf 0}_2\, ,\, {\bf 0}_2)\quad
{\rm clash} \Rightarrow O(7);\nonumber\\
{\rm for}\ \Phi(+\infty) & = & a_{\rm min}\, {\rm diag}({\bf 0}_2\, ,\, \epsilon\, ,\, 
{\bf 0}_2\, ,\, {\bf 0}_2\, ,\, {\bf 0}_2)\quad
{\rm clash} \Rightarrow SO(2)\otimes SO(2) \otimes O(6).
\end{eqnarray}
The $\epsilon$ in the second of these cases can be moved down the diagonal, leading to
differently embedded $SO(2) \otimes SO(2) \otimes O(6)$ invariance groups.

\section{Group theory of kinks from the ${\bf 10}$ and ${\bf 54}$ of $O(10)$}
\label{otherkinks}

We now briefly discuss just the group theoretic aspects of kink
configurations based on the ${\bf 10}$ and ${\bf 54}$ of $O(10)$.

\subsection{Clash of symmetries from the ${\bf 10}$}

Take a single real scalar field $\phi$ in the ${\bf 10}$ of $O(10)$, and denote
it as usual by a $10 \times 1$ column vector.
The global minima induce \cite{li}
\begin{equation}
O(10) \to O(9),
\end{equation}
with
\begin{equation}
\phi_{\rm min} \propto {\rm diag}(0\, ,\, 0\, ,\, 0\, ,\, 0\, ,\, 0\, ,\, 0\, ,\, 
0\, ,\, 0\, ,\, 0\, ,\, 1),
\end{equation}
and permutations thereof. The different permutations correspond
to differently embedded $O(9)$ subgroups. To have a kink displaying the
clash of symmetries, one needs two or more $\phi$ fields with
interchange discrete symmetries between them. If two such fields have
their nonzero entry in different locations, then a kink configuration
can be constructed to induce
\begin{equation}
O(10) \to O(8)
\end{equation}
at all finite values of $z$. This case is very similar to the
$SU(3)$ model studied by Davidson et al.\ \cite{davidson}.

\subsection{Clash of symmetries from the ${\bf 54}$}

The ${\bf 54}$ is the symmetric rank-2 tensor representation. The potential
has a similar form to the ${\bf 45}$, with the possible addition of the
cubic term ${\rm Tr}(\Phi^3)$ since this is nonzero for a symmetric $\Phi$.
From a kink point of view, though, it is interesting to omit this term to
allow the additional discrete symmetry $\Phi \to -\Phi$.

According to Ref.\cite{li}, the global minima are
\begin{equation}
\Phi_{\rm min} \propto {\rm diag}({\bf 0}_9\, ,\, 1)\quad {\rm when}\quad
\lambda_1 >0,\ \lambda_2 < 0,
\end{equation}
and
\begin{equation}
\Phi_{\rm min} \propto {\rm diag}({\bf 1}_5\, ,\, -{\bf 1}_5)\quad {\rm when}\quad
\lambda_1 >0,\ \lambda_2 > 0,
\end{equation}
inducing
\begin{equation}
O(10) \to O(9)\quad {\rm and}\quad O(10) \to O(5) \otimes O(5),
\end{equation}
respectively. As usual, the obvious permutations of these patterns lead
to differently embedded subgroups.

For the $O(10) \to O(9)$ case, the clash of symmetries would further
reduce the symmetry to $O(8)$.

The $O(10) \to O(5) \otimes O(5)$ case can yield clash-induced further
breaking to either $O(4) \otimes O(4)$ or $O(2) \otimes O(2) \otimes
O(3) \otimes O(3)$, depending on how the plus and minus signs are
arranged along the diagonals at $z=\pm\infty$.

\section{Conclusion}
\label{conclusion}

We have discussed several varieties of kinks in $O(10)$ Higgs models,
motivated by the clash of symmetries phenomenon discovered by
Davidson, Toner, Volkas and Wali \cite{davidson}
and independently by Pogosian and
Vachaspati \cite{pg1,pg}. Our focus was primarily on $O(10)$-adjoint kinks that
usually induce the spontaneous breakdown $O(10) \to U(5)$. By carefully
analysing the different ways that $U(5)$ can be embedded in $SO(10)$,
we showed that there are three different kinds of kinks: symmetric,
asymmetric and super-asymmetric. At finite distances from the walls, the invariance
groups are $U(5)$, $U(3) \otimes U(2) = G_{SM} \otimes U(1)'$ 
and $U(4) \otimes U(1)$, respectively.
The last two of these display the clash of symmetries phenomenon.
At the centres of the walls, the unboken symmetries increase to
$O(10)$, $O(6) \otimes U(2)$ and $U(4) \otimes O(2)$, respectively.
Since the connected part of $O(6)$ is $SO(6) = SU(4)$, the asymmetric
kink is connected to both Pati-Salam-like models and the standard model.

All three kinks lie within the same topological class, but they have different
energies. Interestingly, for the special quartic potential that produces
exact hyperbolic-tangent-like kink configurations, the super-asymmetric
kink has the lowest energy and is therefore the topologically stable one.

From the brane-world perspective of Davidson et al., it is encouraging
that the group theory of $O(10)$ allows asymmetric kink solutions,
because the $SO(10) \to SU(4) \otimes SU(2) \otimes U(1) \to G_{SM} \otimes U(1)'$ 
invariance is obviously interesting for model building purposes. 
If our universe is a brane located at the centre of such a wall, then we see the 
$O(10)$ symmetry as being strongly broken to $SU(4) \otimes SU(2) \otimes U(1)$. 
If the brane-world degrees of freedom leak slightly off the
wall, then an additional symmetry breakdown to $U(3) \otimes U(2) = 
G_{SM} \otimes U(1)'$ is effectively
induced on the brane. This provides a possible framework within which to address at
least part of the gauge hierarchy problem: $O(10)$ is strongly broken to 
$SU(4) \otimes SU(2) \otimes U(1)$, then more weakly to $G_{SM} \otimes U(1)'$ depending
on the amount of leakage off the brane.

A realistic brane-world model using this mechanism would require a number of
non-trivial additional features, such as the inclusion of gravity, a rationale for 
why some fields are confined to the brane while others propagate into the extra
dimension, and a suitable symmetry breakdown pattern induced by a topologically
stable domain wall. Our results are also of general interest in the continuing
study of kink and other soliton solutions of non-trivial field theories.

\acknowledgments{We thank L. Pogosian, A. Davidson and K. Wali 
for helpful correspondence.
RRV also thanks A. Demaria and C. Low for useful discussions.
This work was supported in part by the Australian
Research Council.}

\appendix

\section{Deriving $O(10) \to U(3) \otimes U(2)$ for asymmetric kinks}

We now prove that asymmetric kinks are invariant under $U(3) \otimes U(2)$.
Similar reasoning shows that super-asymmetric kinks are invariant under
$U(4) \otimes U(1)$.

The asymmetric kink configuration is of the form
\begin{equation}
\Phi(z) = \alpha(z)\, \Phi(-\infty) + \beta(z)\, \Phi(+\infty),
\end{equation}
where $\Phi(-\infty)$ is given by Eq.\ (\ref{eq:minusbc}), $\Phi(+\infty)$
is given by the middle line of Eq.\ (\ref{eq:plusbc}), $a_{\rm min}\alpha = -(f+g)/2$
and $a_{\rm min}\beta = (f-g)/2$. We seek the invariance group at any $z$ obeying
$0 < |z| < \infty$. Consider $SO(10)$ transforms $\Phi \to A \Phi A^T$.
By linearity,
\begin{equation}
\Phi \to \alpha\, A \Phi(-\infty) A^T + \beta\, A \Phi(+\infty) A^T.
\end{equation}
At the unexceptional points $0 < |z| < \infty$, the invariance group
of $\Phi(z)$ is given by the set of matrices $A$ that simultaneously
leave $\Phi(-\infty)$ and $\Phi(+\infty)$ invariant. This is the
intersection of the $U(5)_h$ and $U(5)_1$ groups defined in the
main text [see Eq.\ (\ref{eq:U5one})]. 

Now, a general $U(5)_h$ matrix $A^{(h)}$ has the form
\begin{equation}
A^{(h)} = \left( \begin{array}{ccccc}
c_{11} & c_{12} & c_{13} & c_{14} & c_{15} \\
c_{21} & c_{22} & c_{23} & c_{24} & c_{25} \\
c_{31} & c_{32} & c_{33} & c_{34} & c_{35} \\
c_{41} & c_{42} & c_{43} & c_{44} & c_{45} \\
c_{51} & c_{52} & c_{53} & c_{54} & c_{55}
\end{array} \right),
\end{equation}
where each entry is of the form
\begin{equation}
c_{ij} = r_{ij} \left(
\begin{array}{cc}
\cos\theta_{ij} & -\sin\theta_{ij} \\
\sin\theta_{ij} & \cos\theta_{ij}
\end{array} \right)
\end{equation}
with real $r_{ij}$. A general $U(5)_1$ matrix $A^{(1)}$
is of
the form $Q_1 A^{(h)} Q_1^T$, as per Eq.\ (\ref{eq:U5one}).
Direct matrix multiplication shows that
\begin{equation}
A^{(1)} = \left( \begin{array}{ccccc}
c_{11} & c_{12} & c_{13} & d_{14} & d_{15} \\
c_{21} & c_{22} & c_{23} & d_{24} & d_{25} \\
c_{31} & c_{32} & c_{33} & d_{34} & d_{35} \\
d_{41} & d_{42} & d_{43} & c_{44}^* & c_{45}^* \\
d_{51} & d_{52} & d_{53} & c_{54}^* & c_{55}^*
\end{array} \right),
\end{equation}
where 
\begin{equation}
c_{ij}^* \equiv r_{ij}
\left(
\begin{array}{cc}
\cos(-\theta_{ij}) & -\sin(-\theta_{ij}) \\
\sin(-\theta_{ij}) & \cos(-\theta_{ij})
\end{array} \right),
\end{equation}
and
\begin{equation}
d_{ij} \equiv r_{ij}
\left(
\begin{array}{cc}
\pm\sin\theta_{ij} & \cos\theta_{ij} \\
\cos\theta_{ij} & \mp\sin\theta_{ij}
\end{array} \right).
\end{equation}
The upper signs come from left-multiplication by
$\sigma_1$, while the lower signs come from right multiplication.
We now observe that while the set of $c_{ij}$'s
is the same as the set of $c_{ij}^*$'s, almost no
matrix of $d_{ij}$ form can ever have $c_{ij}$ form because
$\det(d_{ij}) \le 0$ while $\det(c_{ij}) \ge 0$. The only
overlap is for $r_{ij}=0$. The intersection of $U(5)_h$ and
$U(5)_1$ thus consists of $10 \times 10$ matrices with
nonzero $6 \times 6$ and $4 \times 4$ matrices along the diagonal
and zeroes everywhere else. These two blocks lead to independent
$U(3)$ and $U(2)$ invariances, as the $h$-map immediately reveals.

\end{document}